\documentclass[letterpaper]{article} 
\usepackage{aaai24} 
\usepackage{times}  
\usepackage{helvet}  
\usepackage{courier}  
\usepackage[hyphens]{url}  
\usepackage{graphicx} 
\usepackage{natbib}  
\usepackage{amsmath}
\usepackage{caption} 
\usepackage{algorithm}
\usepackage{algorithmic}

\usepackage{newfloat}
\usepackage{listings}
\DeclareCaptionStyle{ruled}{labelfont=normalfont,labelsep=colon,strut=off} 
\floatstyle{ruled}
\newfloat{listing}{tb}{lst}{}
\floatname{listing}{Listing}
\usepackage{xcolor}
\usepackage{amssymb}

\usepackage{csquotes}

\usepackage{tikz}
\usepackage{pgfplots}
\usepackage{MnSymbol}
\usepackage{subcaption}

\usetikzlibrary{intersections}
\usetikzlibrary{patterns}
\pgfplotsset{compat=1.14, set layers}

\definecolor{codegreen}{rgb}{0,0.6,0}
\definecolor{codegray}{rgb}{0.5,0.5,0.5}
\definecolor{codepurple}{rgb}{0.58,0,0.82}
\definecolor{backcolour}{rgb}{0.95,0.95,0.92}

\lstdefinestyle{mystyle}{
    basicstyle=\ttfamily\scriptsize,
    backgroundcolor=\color{backcolour},   
    commentstyle=\color{codegreen},
    keywordstyle=\color{magenta},
    stringstyle=\color{codepurple},
    breakatwhitespace=false,         
    breaklines=true,                 
    captionpos=b,                    
    keepspaces=false,                 
    numbers=none,                    
    numbersep=3pt,                  
    showspaces=false,                
    showstringspaces=false,
    showtabs=false,                  
    tabsize=2
}

\usepackage{booktabs}
\usepackage{multirow}

\title{Diagnosing Infeasible Optimization Problems Using Large Language Models}
\author {
    Hao Chen\textsuperscript{\rm 1},
    Gonzalo E. Constante-Flores\textsuperscript{\rm 1},
    Can Li\textsuperscript{\rm 1}
}
\affiliations {
    \textsuperscript{\rm 1}Purdue University\\
\{chen4433, geconsta, canli\}@purdue.edu
}

\begin{document}

\maketitle

\begin{abstract}
Decision-making problems can be represented as mathematical optimization models, finding wide applications in fields such as economics, engineering and manufacturing, transportation, and health care. Optimization models are mathematical abstractions of the problem of making the best decision while satisfying a set of requirements or constraints. One of the primary barriers to deploying these models in practice is the challenge of helping practitioners understand and interpret such models, particularly when they are infeasible, meaning no decision satisfies all the constraints. Existing methods for diagnosing infeasible optimization models often rely on expert systems, necessitating significant background knowledge in optimization. In this paper, we introduce OptiChat, a first-of-its-kind natural language-based system equipped with a chatbot GUI for engaging in interactive conversations about infeasible optimization models. OptiChat can provide natural language descriptions of the optimization model itself, identify potential sources of infeasibility, and offer suggestions to make the model feasible. The implementation of OptiChat is built on GPT-4, which interfaces with an optimization solver to identify the minimal subset of constraints that render the entire optimization problem infeasible, also known as the Irreducible Infeasible Subset (IIS). We utilize few-shot learning, expert chain-of-thought, key-retrieve, and sentiment prompts to enhance OptiChat's reliability.
Our experiments demonstrate that OptiChat assists both expert and non-expert users in improving their understanding of the optimization models, enabling them to quickly identify the sources of infeasibility.
\end{abstract}
\begin{figure}[t]
    \centering
    \includegraphics[trim={5cm 0 5.5cm 9cm},clip, width=\linewidth]{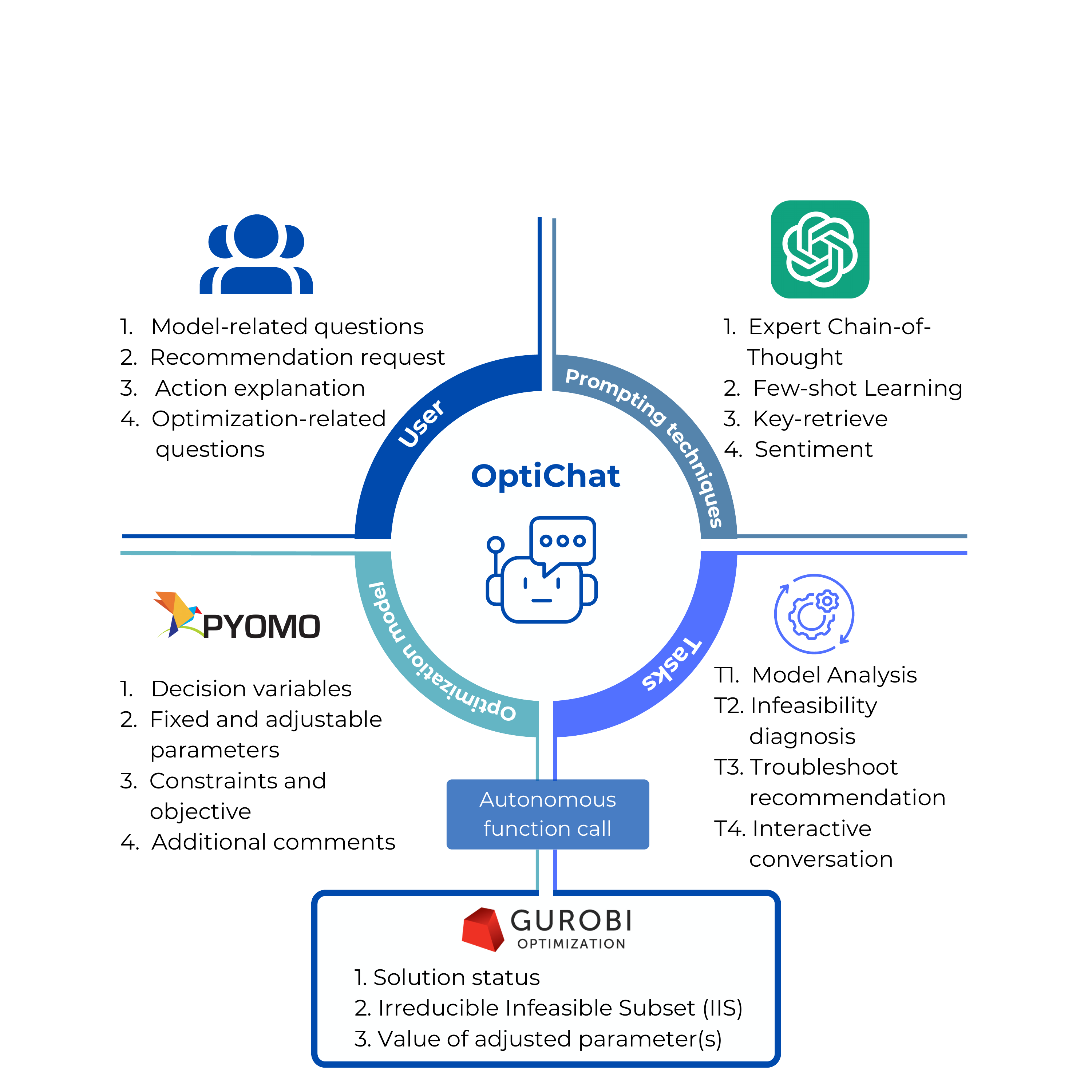}
    \caption{\textbf{Overview of OptiChat.} Users can directly interact with OptiChat using natural language without writing a single line of code. OptiChat is an LLM-powered autonomous agent system.}
    \label{fig:overview}
\end{figure}
\section{Introduction}
Mathematical optimization has wide applications in real-world decision-making problems such as aircraft crew scheduling, smart grid operation, and portfolio optimization \cite{Rardin2016}. Most of these applications can be formulated as a mixed-integer linear program (MILP):
\begin{equation}\label{eq:MILP}
\begin{aligned}
  \min_{\mathbf{x}} && \mathbf{c}^\top \mathbf{x}\\
  \text{subject to } && \mathbf{A} \mathbf{x} &\le \mathbf{b}\text{,}\\
  && \mathbf{x} &\in \mathbb{Z}^{p} \times \mathbb{R}^{n-p} \text{.}
\end{aligned}
\end{equation}
where decisions to make are denoted as $\mathbf{x}$ that consist of both integer and continuous variables; $\mathbf{c}$ represents the cost coefficients; the problem is subject to linear constraints $\mathbf{A} \mathbf{x} \le \mathbf{b}$: $\mathbf{A} \in \mathbb{R}^{m \times n}$,  $\mathbf{b} \in \mathbb{R}^m$. An infeasible \eqref{eq:MILP} means that there exists no solution $\mathbf{x} \in \mathbb{Z}^{p} \times \mathbb{R}^{n-p}$ that satisfies all the constraints.

\begin{figure}[ht]
     \centering
     \begin{subfigure}[b]{0.23\textwidth}
         \centering
         \scalebox{0.5}{
         \begin{tikzpicture}[E/.style={font=\LARGE,text=black, sloped, pos=0.75,framed}]
            \begin{axis}[
                axis line style=thick, 
                axis lines=center,
                axis on top,
                xmin=-5, xmax=7,
                ymin=-7, ymax=7,
                axis line style={draw=none},
                tick style={draw=none},
                xticklabels={},yticklabels={}
                every axis plot post/.append style={very thick, color=blue!50}
            ]
            \draw (-5,-7) -- (-5,7) -- (7,7) -- (7,-7) -- (-5,-7);
            \newcommand{\drawge}{-- (rel axis cs:1,0) -- (rel axis cs:1,1) -- (rel axis cs:0,1) \closedcycle}
            \newcommand{\drawle}{-- (rel axis cs:1,1) -- (rel axis cs:1,0) -- (rel axis cs:0,0) \closedcycle}
            
            \addplot[color= red!75,name path=A,domain=-7:7, pattern=north west lines, pattern color=red!60!white, draw=red!60!white] {x-0.1} \drawge;
            \addplot[color=green!60!black, name path=B,domain=-7:7, pattern=north west lines, pattern color=green!60!black, draw=green!60!black] {0.2*x-1} \drawle;
            \addplot[color=blue!75, name path=C,domain=-7:7, pattern=north west lines, pattern color=blue!60!white, draw=blue!60!white] {-x+2} \drawge;
            
            \draw[thick, color=red!75]
                (-7,-7.1) -- (7, 6.9) node[pos=0.9,above]{$A$}; 
            \draw[thick, color=green!60!black]
                (-7,-2.4) -- (7, 0.4) node[pos=0.25,above]{$B$}; 
            \draw[thick, color=blue!75]
                (-7,9) -- (7, -5) node[pos=0.9,below]{$C$}; 
            \draw (-5,-7) -- (-5,7) -- (7,7) -- (7,-7) -- (-5,-7);
            \end{axis}
        \end{tikzpicture}}
         \caption{Original constraint set}
         \label{fig:iis_a}
     \end{subfigure}
     \begin{subfigure}[b]{0.23\textwidth}
         \centering
         \scalebox{0.5}{
         \begin{tikzpicture}[E/.style={font=\LARGE,text=black, sloped, pos=0.75,framed}]
            \begin{axis}[
                axis line style=thick, 
                axis lines=center,
                axis on top,
                xmin=-5, xmax=7,
                ymin=-7, ymax=7,
                axis line style={draw=none},
                tick style={draw=none},
                xticklabels={},yticklabels={}
                every axis plot post/.append style={very thick, color=blue!50}
            ]
            \newcommand{\drawge}{-- (rel axis cs:1,0) -- (rel axis cs:1,1) -- (rel axis cs:0,1) \closedcycle}
            \newcommand{\drawle}{-- (rel axis cs:1,1) -- (rel axis cs:1,0) -- (rel axis cs:0,0) \closedcycle}
            
            \addplot[color=green!60!black, name path=B,domain=-7:7, pattern=north west lines, pattern color=green!60!black, draw=green!60!black] {0.2*x-1} \drawle;
            \addplot[color=blue!75, name path=C,domain=-7:7, pattern=north west lines, pattern color=blue!60!white, draw=blue!60!white] {-x+2} \drawge;

            \fill [][very thick, pattern=crosshatch, pattern color=black!80!white](2.5,-0.5)--(7,0.4)--(7,-5)--(2.5,-0.5);

            \draw[thick, color=green!60!black]
                (-7,-2.4) -- (7, 0.4) node[pos=0.25,above]{$B$}; 
            \draw[thick, color=blue!75]
                (-7,9) -- (7, -5) node[pos=0.25,below]{$C$}; 
            \draw (-5,-7) -- (-5,7) -- (7,7) -- (7,-7) -- (-5,-7);
            \end{axis}
        \end{tikzpicture}}
         \caption{Remove constraint A}
         \label{fig:iis_b}
     \end{subfigure}
     \hfill
     \begin{subfigure}[b]{0.23\textwidth}
         \centering
         \scalebox{0.5}{
         \begin{tikzpicture}[E/.style={font=\LARGE,text=black, sloped, pos=0.75,framed}]
            \begin{axis}[
                axis line style=thick, 
                axis lines=center,
                axis on top,
                xmin=-5, xmax=7,
                ymin=-7, ymax=7,
                axis line style={draw=none},
                tick style={draw=none},
                xticklabels={},yticklabels={}
                every axis plot post/.append style={very thick, color=blue!50}
            ]
            \newcommand{\drawge}{-- (rel axis cs:1,0) -- (rel axis cs:1,1) -- (rel axis cs:0,1) \closedcycle}
            \newcommand{\drawle}{-- (rel axis cs:1,1) -- (rel axis cs:1,0) -- (rel axis cs:0,0) \closedcycle}
            
            \addplot[color= red!75,name path=A,domain=-7:7, pattern=north west lines, pattern color=red!60!white, draw=red!60!white] {x-0.1} \drawge;
            \addplot[color=blue!75, name path=C,domain=-7:7, pattern=north west lines, pattern color=blue!60!white, draw=blue!60!white] {-x+2} \drawge;

            \fill [][very thick, pattern=crosshatch, pattern color=black!80!white](1.05,0.95)--(7.1,7)--(-5,7)--(1.05,0.95);
            
            \draw[thick, color=red!75]
                (-7,-7.1) -- (7, 6.9) node[pos=0.25,below]{$A$}; 
            \draw[thick, color=blue!75]
                (-7,9) -- (7, -5) node[pos=0.9,below]{$C$}; 
            \draw (-5,-7) -- (-5,7) -- (7,7) -- (7,-7) -- (-5,-7);
            \end{axis}
        \end{tikzpicture}}
         \caption{Remove constraint B}
         \label{fig:iis_c}
     \end{subfigure}
     \begin{subfigure}[b]{0.23\textwidth}
         \centering
         \scalebox{0.5}{
         \begin{tikzpicture}[E/.style={font=\LARGE,text=black, sloped, pos=0.75,framed}]
            \begin{axis}[
                axis line style=thick, 
                axis lines=center,
                axis on top,
                xmin=-5, xmax=7,
                ymin=-7, ymax=7,
                axis line style={draw=none},
                tick style={draw=none},
                xticklabels={},yticklabels={}
                every axis plot post/.append style={very thick, color=blue!50}
            ]
            
            \newcommand{\drawge}{-- (rel axis cs:1,0) -- (rel axis cs:1,1) -- (rel axis cs:0,1) \closedcycle}
            \newcommand{\drawle}{-- (rel axis cs:1,1) -- (rel axis cs:1,0) -- (rel axis cs:0,0) \closedcycle}
            
            \addplot[color= red!75,name path=A,domain=-7:7, pattern=north west lines, pattern color=red!60!white, draw=red!60!white] {x-0.1} \drawge;
            \addplot[color=green!60!black, name path=B,domain=-7:7, pattern=north west lines, pattern color=green!60!black, draw=green!60!black] {0.2*x-1} \drawle;

            \fill [][very thick, pattern=crosshatch, pattern color=black!80!white](-1.125,-1.225)--(-7,-7.1)--(-7,-2.4)--(-1.125,-1.225);
            
            \draw[thick, color=red!75]
                (-7,-7.1) -- (7, 6.9) node[pos=0.9,below]{$A$}; 
            \draw[thick, color=green!60!black]
                (-7,-2.4) -- (7, 0.4) node[pos=0.9,above]{$B$}; 
            \draw (-5,-7) -- (-5,7) -- (7,7) -- (7,-7) -- (-5,-7);
            \end{axis}
        \end{tikzpicture}}
         \caption{Remove constraint C}
         \label{fig:iis_d}
     \end{subfigure}
     \caption{Example of Irreducible Infeasible Subset (IIS). The feasible region of each constraint is colored. The original constraint set, $S=\{A,B,C\}$ is infeasible since no solution satisfies the three constraints. However, any proper subset is a feasible constraint set, i.e., if any of the constraints are dropped, then the constraint set becomes feasible.}
     \label{fig:iis}
\end{figure}
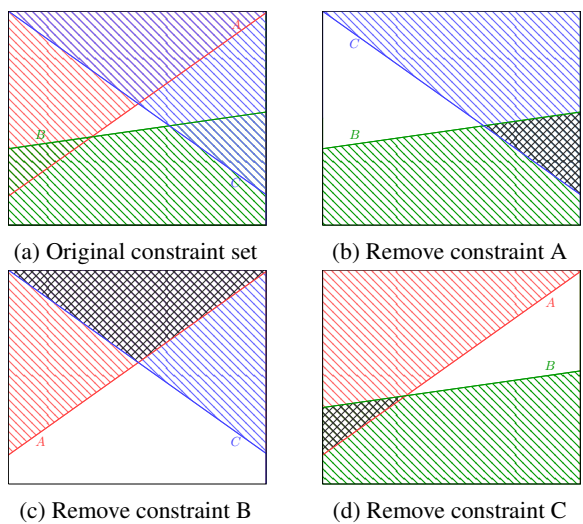

The optimization community has come up with different mathematical concepts to characterize infeasible optimization models: the most commonly used one is the Irreducible Infeasible Subset (IIS) \cite{Chinneck1991}, which refers to a minimal subset of constraints and/or variable bounds within an optimization model that is infeasible. An ISS is a subset of an infeasible model with two properties: (i) the ISS is infeasible, and (ii) any proper subset of the ISS is feasible. (see Figure \ref{fig:iis} for an illustrative example). By isolating the core conflicting constraints, IIS can help analyze and diagnose infeasible models.

Optimization experts can gain insight into what causes the infeasibility from the IIS. However, practitioners without much knowledge of mathematical programming often treat these models as black boxes. Therefore, when practitioners encounter an infeasible optimization model, their ability to find the root causes and take corrective actions is limited. To resolve this, a Human-Computer Interaction (HCI) system is needed to help humans diagnose infeasible optimization models.

One of the pioneering efforts to develop an HCI tool for analyzing infeasible optimization models is the ANALYZE system \cite{Greenberg1983,Greenberg1987} developed in the 1980s, an expert system that relies on a specific syntax tailored for optimization problems. The downside of expert systems is that they still require a fair amount of domain knowledge and optimization background. In the pioneering paper \cite{Greenberg1983}, the author insightfully foresaw that ``Ideally, one would like to move toward a natural language query system where the descriptions are resident with the model in a way that permits more automation to obtain answers.'' However, to the best of our knowledge, such a natural language-based system envisioned by Greenberg has not been developed due to the lack of a general-purpose language model. Recently, large language models (LLMs) such as GPT-4 \cite{OpenAI2023} and LLaMA \cite{Touvron2023} have achieved remarkable success in diverse applications, opening new ways of developing software for analyzing optimization models.


In this paper, we develop OptiChat, a chatbot for providing natural language explanations of infeasible optimization models and receiving questions and feedback from human users, taking advantage of the GPT-4 APIs. To this end, OptiChat interacts with an optimization solver to identify the IIS and modify the model parameters by adding slack variables to make the model feasible based on human feedback. We use state-of-the-art prompting techniques, such as few-shot learning, expert chain-of-thought, and sentiment prompting to make the process robust. Additionally, we propose a \textit{key-retrieve} prompting technique to improve OptiChat's robustness further.

\section{Related Works}
\subsection{Algorithms for Isolating IIS Constraints}
A simple algorithm to isolate an IIS  is the \textit{deletion filter} \cite{Chinneck1991} where each constraint is first tentatively dropped from the constraint set; if the problem becomes feasible, then the constraint will be returned to the constraint set; otherwise, the constraint is kept permanently. A single iteration over all the constraints is guaranteed to find an IIS. Other IIS isolation methods include the additive method \cite{Tamiz1996} and the combination of the deletion filter and the additive method \cite{Guieu1999}. IIS isolation algorithms have been implemented in commercial optimization solvers such as CPLEX \cite{CPLEX2022}, Gurobi \cite{Gurobi}, and Mosek \cite[\S~14.2]{Mosek}. Infeasible linear programs can be repaired by adding slack variables to each constraint and penalizing the slacks in the objective function. A complete review of IIS detection can be found in the monograph \cite{Chinneck2008}. 

\subsection{Expert Systems for Optimization Models}
Expert systems, such as ANALYZE \cite{Greenberg1983, Greenberg1987, Greenberg1993,Greenberg1993IS}, emerged in the 1980s to analyze optimization models. ANALYZE is equipped to diagnose infeasibility, conduct sensitivity analysis, and generate views for linear programming problems. Commercial software such as AMPL \cite{Fourer1989}, AIMMS \cite{Bisschop2004}, and GAMS \cite{Bussieck2004} offer algebraic modeling languages specifically tailored for optimization problems. These modeling platforms also provide capabilities for the analysis and visualization of results. In recent years, open-source modeling libraries like Pyomo \cite{Hart2011} and JuMP \cite{Lubin2023} have been introduced, leveraging the Python and Julia programming languages, respectively. A notable advantage of these open-source libraries is their flexibility, allowing programmers and optimization experts to integrate them smoothly with other libraries. However, a significant limitation of these expert systems is their reliance on programming-language-like syntax. This necessitates a robust background in both optimization and programming for users to effectively interpret optimization models and their results.

\begin{figure*}[h!]
    \centering
    \includegraphics[width=0.8\textwidth]{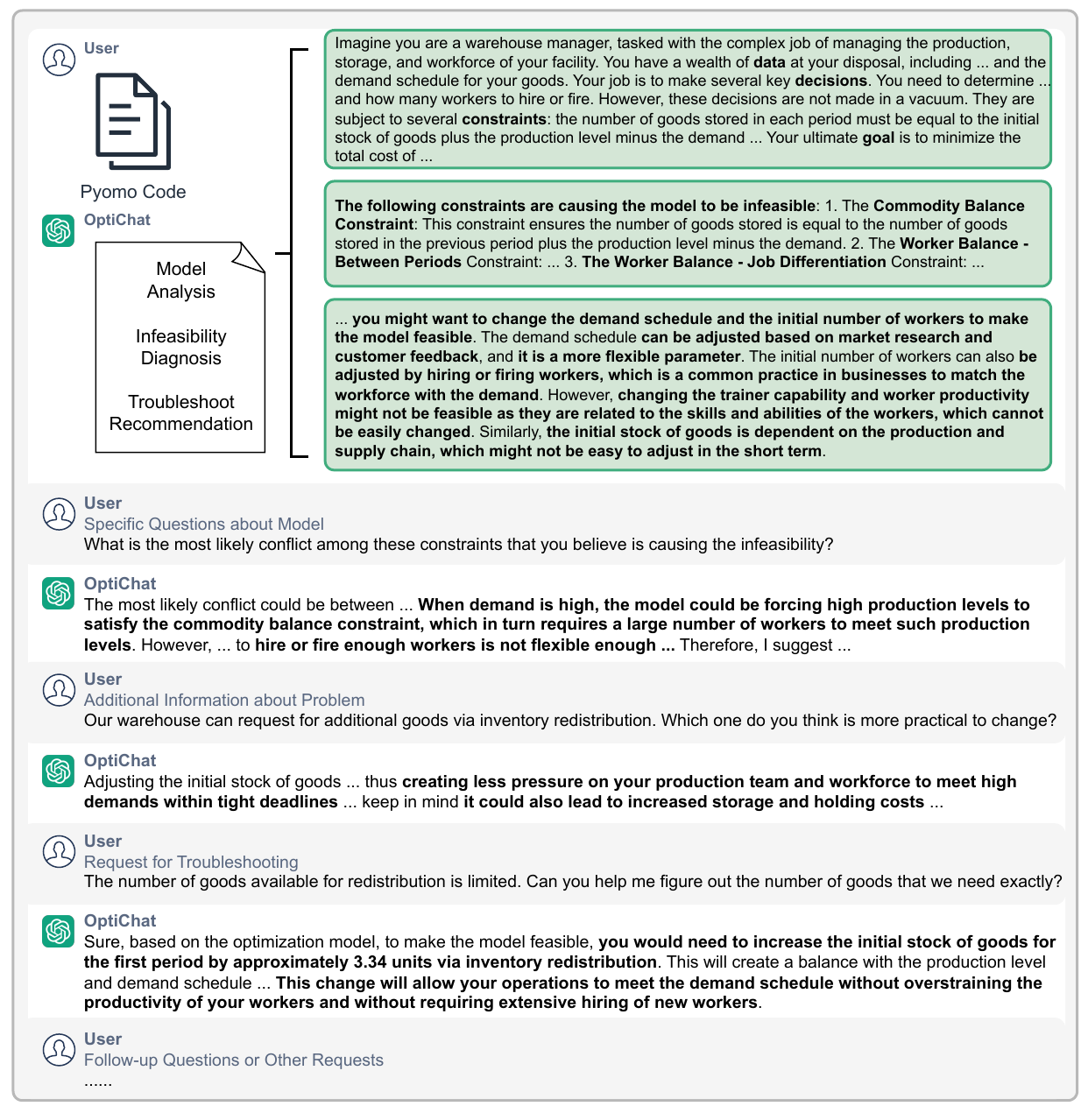}
    \caption{\textbf{Illustrative conversation.} This example shows how the OptiChat agent conducts model analysis, infeasibility diagnosis, troubleshoot recommendations, and interactive conversations through the GUI. }
    \label{fig:conversationexample}
\end{figure*}
\subsection{Natural Language Dialogue Systems for Machine Learning Models}
To the best of our knowledge, no natural language systems exist for optimization models. However, a related topic is the natural language dialogue systems for explaining machine learning models.  The prototype design of these interactive dialogue systems is typically equipped with functionalities that explain the potential impact of input modification on output \cite{Sokol2018,Wexler2020}, which reflects the model's decision-making in different situations. Converting these explanations into natural language dialogues is more effective in convincing users \cite{Feldhus2022}. Previous designs only address questions within specific contexts and lack flexibility. TalkToModel \cite{Slack2023} is a recently developed chatbot for explaining ML models by combining LLMs with external tools such as counterfactual explanations. 

\subsection{Prompt Engineering}
For the LLM to operate with high accuracy, crafting the right prompts is crucial. Prompt Engineering encompasses methods that direct the behavior of LLM towards specific results without altering its weight parameters. In zero-shot learning, one simply provides the task description to the model and awaits the output. Few-shot learning \cite{zhao2021calibrate} introduces a set of exemplary demonstrations, each comprising both the input and the anticipated output for the target task. Chain-of-thought (CoT) prompting \cite{wei2022chain} produces a series of concise sentences, each detailing a step in the reasoning process, eventually resulting in the final answer.  For an updated review on prompt engineering, we refer to the blog \cite{weng2023prompt}.

\section{Method}
OptiChat is an interactive dialogue system that facilitates seamless troubleshooting through natural language-based questions and requests by non-experts. As shown in Figure \ref{fig:overview},  it is an agent-based autonomous system that coordinates the user, the optimization model,  the optimization solver, and the LLM. The non-expert user is interested in understanding the optimization model written in an algebraic modeling language, in this case, the Pyomo/Python modeling framework. The Pyomo model has symbolic expressions of the decision variables, the input parameters, the constraints, and the objective.  The optimization solver can be used to solve the Pyomo model; obtain the solution status; retrieve the IIS if the model becomes infeasible; and solve the model with adjusted parameters to make it feasible. Most importantly, the LLM functions as the brain of the agent: it processes the optimization model to provide the user with natural language descriptions; engages in conversations with the user; understands which parameters should be changed, and decides when to use the optimization solver based on the conversations. Given the functions of these agents, OptiChat can accomplish four major tasks: provide the user with a contextual model description; analyze the potential sources of infeasibility, give recommendations for making the model feasible, and have interactive conversations with the user to answer follow-up questions. In what follows, we first provide an illustrative example of using OptiChat to showcase a typical workflow of infeasibility diagnosis.  Second, we describe the technical details of how the optimization solver is used to isolate the IIS and make the model feasible. Third, we explain how the LLM coordinates the user input, the optimization model, and the solver. 
\subsection{Illustrative Conversations with OptiChat}
Before delving into the technicalities, we present an illustrative example of the \textit{on-the-job training} problem, adapted from the GAMS modeling library \footnote{\url{https://www.gams.com/latest/gamslib_ml/libhtml/index.html}}. This example, showcased in Figure \ref{fig:conversationexample}, demonstrates user interaction with OptiChat. Initially, OptiChat receives an infeasible Pyomo instance related to the on-the-job training problem. Subsequently, it offers the user a clear overview of the problem. Causes of infeasibility are diagnosed and conveyed to the user in easily understandable language. Moreover, OptiChat suggests measures to regain feasibility. After this initial analysis, users are free to pose further questions about the model. For instance, they can inquire about key conflicts within the on-the-job training problem or even request changes to the model's data to find feasible solutions.

\subsection{Usage of the Optimization Model and Solver}
The LLM itself cannot provide an end-to-end solution for infeasibility diagnosis. The two other agents in Figure \ref{fig:overview}, the optimization model and the optimization solver are indispensable for OptiChat to have a robust performance.  In this subsection, we explain how the optimization model and solver are used for identifying the IIS and recovering the feasibility. To this end, we first illustrate how an optimization expert would diagnose an infeasible optimization model, which will be used as the basis for prompting the LLM. 
Optimization experts typically begin with the code or mathematical formulation to understand the underlying problem represented by the model. To identify the causes of infeasibility, experts typically compute the IIS of the model to recognize the main conflicts, which allows them to extract valuable insights from a subset of constraints. Next, to resolve the infeasibility, optimization experts either remove some of the constraints from IIS or add slack variables. 

\subsubsection{Computing IIS} 
We use the optimization solver, Gurobi, to compute the IIS. Gurobi uses different algorithms for computing IIS for linear programs and mixed-integer linear programs. For linear programming problems, i.e., $\mathbf{x}\in  \mathbb{R}^{n}$,  it was shown in \cite{gleeson1990identifying} that  the support (indices corresponding to the nonzero values) of any vertex of the following polyhedral set 
\begin{equation}
    \label{IIS}
    P = \{\mathbf{y} \in  \mathbb{R}^{m}| \mathbf{y}^{\top} \mathbf{A} = 0, \mathbf{y}^{\top} \mathbf{b} \leq -1, \mathbf{y} \geq 0\}
\end{equation}
gives an IIS. Therefore, enumerating over all the IIS is equivalent to enumerating to the vertices of $P$. This claim can be proved using the Farkas' lemma \cite{gleeson1990identifying}. 

For mixed-integer linear programs, the  Farkas' lemma argument is no longer applicable. In this case, Gurobi uses logic-based methods such as deletion filter and additive method reviewed in the previous section.

\subsubsection{Resolving Infeasibility} 
\begin{figure}[ht]
     \centering
     \begin{subfigure}[b]{0.23\textwidth}
         \centering
         \scalebox{0.5}{
         \begin{tikzpicture}[E/.style={font=\LARGE,text=black, sloped, pos=0.75}]
            \begin{axis}[
                axis line style=thick, 
                axis lines=center,
                axis on top,
                xmin=-5, xmax=7,
                ymin=-7, ymax=7,
                axis line style={draw=none},
                tick style={draw=none},
                xticklabels={},yticklabels={}
                every axis plot post/.append style={very thick, color=blue!50}
            ]
            \newcommand{\drawge}{-- (rel axis cs:1,0) -- (rel axis cs:1,1) -- (rel axis cs:0,1) \closedcycle}
            \newcommand{\drawle}{-- (rel axis cs:1,1) -- (rel axis cs:1,0) -- (rel axis cs:0,0) \closedcycle}
            
            \addplot[color= red!75,name path=A,domain=-7:7, pattern=north west lines, pattern color=red!60!white, draw=red!60!white] {-0.15*x-0.1} \drawge;
            \addplot[color=green!60!black, name path=B,domain=-7:7, pattern=north west lines, pattern color=green!60!black, draw=green!60!black] {0.2*x-1} \drawle;
            \addplot[color=blue!75, name path=C,domain=-7:7, pattern=north west lines, pattern color=blue!60!white, draw=blue!60!white] {-x+2} \drawge;

            \fill [name intersections={of=A and B,by={2.5714286,0.414286}}, name intersections={of=A and C,by={2.235294,-0.235294}}, name intersections={of=B and C,by={2.5,-0.5}}][very thick, pattern=crosshatch, pattern color=black!80!white](2.5,-0.5)--(7,0.4)--(7,-1.15)--(7,-1.15);
            
            \draw[thick, color=red!75]
                (-7,0.95) -- (7, -1.15) node[pos=0.25,above]{$A$}; 
            \draw[thick, color=green!60!black]
                (-7,-2.4) -- (7, 0.4) node[pos=0.25,above]{$B$}; 
            \draw[thick, color=blue!75]
                (-7,9) -- (7, -5) node[pos=0.9,below]{$C$}; 
            
            \end{axis}
        \end{tikzpicture}}
         \caption{Rotate constraint A}
         \label{fig:feas_a}
     \end{subfigure}
     \begin{subfigure}[b]{0.23\textwidth}
         \centering
         \scalebox{0.5}{
         \begin{tikzpicture}[E/.style={font=\LARGE,text=black, sloped, pos=0.75}]
            \begin{axis}[
                axis line style=thick, 
                axis lines=center,
                axis on top,
                xmin=-5, xmax=7,
                ymin=-7, ymax=7,
                axis line style={draw=none},
                tick style={draw=none},
                xticklabels={},yticklabels={}
                every axis plot post/.append style={very thick, color=blue!50}
            ]
            \newcommand{\drawge}{-- (rel axis cs:1,0) -- (rel axis cs:1,1) -- (rel axis cs:0,1) \closedcycle}
            \newcommand{\drawle}{-- (rel axis cs:1,1) -- (rel axis cs:1,0) -- (rel axis cs:0,0) \closedcycle}

            \addplot[color= red!75,name path=A,domain=-7:7, pattern=north west lines, pattern color=red!60!white, draw=red!60!white] {x-3} \drawge;
            \addplot[color=green!60!black, name path=B,domain=-7:7, pattern=north west lines, pattern color=green!60!black, draw=green!60!black] {0.2*x-1} \drawle;
            \addplot[color=blue!75, name path=C,domain=-7:7, pattern=north west lines, pattern color=blue!60!white, draw=blue!60!white] {-x+2} \drawge;
            
            \draw[thick, color=red!75]
                (-7,-10) -- (7, 4) node[pos=0.95,above]{$A$}; 
            \draw[thick, color=blue!75]
                (-7,9) -- (7, -5) node[pos=0.9,below]{$C$}; 
            \draw[thick, color=green!60!black]
                (-7,-2.4) -- (7, 0.4) node[pos=0.25,above]{$B$}; 
            \filldraw [black!80!white] (2.5,-0.5) circle (2pt);
            \end{axis}
        \end{tikzpicture}}
         \caption{Translate constraint A }
         \label{fig:feas_b}
     \end{subfigure}
    \caption{\textbf{Effects of adding slack variables.} (a) shows the problem can be made feasible by adjusting the left-hand side matrix $\mathbf{A}$. (b) shows the effect of adding slack variable $\mathbf{\delta b}^+$ to the right-hand side. The region and the point in black represent the feasible regions after adding the slack variables.}
    \label{fig:feasible_set}
\end{figure}
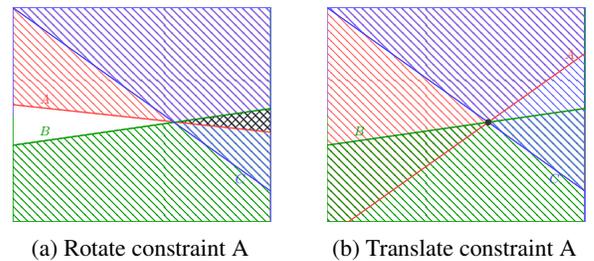

Optimization experts have two different ways of restoring feasibility: (1) remove the constraints detected by the IIS algorithms recursively until the model becomes feasible; (2) add slack variables to the optimization problems by changing the input parameters. OptiChat takes the second approach because removing the constraints may not be practical in real-world applications. On the other hand, the slack-variable approach can give the decision-maker a concrete actionable plan for making the model feasible. More precisely, we modify the parameters in the matrix $\mathbf{A}$ and the right-hand side vector $\mathbf{b}$ in the original MILP \eqref{eq:MILP} by adding slack variables. Geometric interpretations of adding the slack variables are shown in Figure \ref{fig:feasible_set}.  

Mathematically, the following extended problem is solved,
\begin{equation}\label{eq:MILPwithSlack}
\begin{aligned}
  \min_{\mathbf{x, \delta A^+, \delta A^-, \delta b^+}} & \quad \sum_{(i,j)\in \mathcal{S}_{A}} \big(\mathbf{\delta A}_{ij}^+ + \mathbf{\delta A}_{ij}^- \big)+ \sum_{i\in\mathcal{S}_{b}}\mathbf{\delta b}_{i}^+ \\
  \text{subject to } & (\mathbf{A} + \mathbf{\delta A^+} - \mathbf{\delta A^-}) \mathbf{x} \le \mathbf{b} + \mathbf{\delta b^+} \text{,}\\
  & \mathbf{x} \in \mathbb{Z}^{p} \times \mathbb{R}^{n-p},\quad  \mathbf{\delta A^+, \delta A^-, \delta b^+}\geq 0\text{.}
\end{aligned}
\end{equation}
where $\mathbf{\delta A^+}$, $\mathbf{\delta A^-}$, $\mathbf{\delta b^+}$ represent the slack variables. All the slack variables added are nonnegative. Since we only have inequality constraints, adding nonnegative slacks on the right-hand side relaxes the problem. For the left-hand side matrix, we have both $ \mathbf{\delta A^+} $ and $\mathbf{\delta A^-}$. Note that the slack variables have different dimensionality compared with the parameters $\mathbf{A}$ and $\mathbf{b}$. This is because only a subset of the parameters can be adjusted in practice.  Determining which parameters to change is based on the judgment of the LLM and the request of the user, which will be discussed in the next subsection. The support sets of the slack variables are denoted as $\mathcal{S}_{A}$ and $\mathcal{S}_{b}$, respectively. The objective is to minimize the total perturbation to the original problem by adding up all the slack variables. In principle, we could assign different weights to different slack variables representing the cost associated with changing the corresponding parameters. 

When the support set $\mathcal{S}_{A}$ is empty, i.e., only the right-hand side parameters are perturbed, problem \eqref{eq:MILPwithSlack} remains to be a mixed-integer linear program (MILP), which can be solved as fast as the original problem \eqref{eq:MILP}.  However, when set $\mathcal{S}_{A}$ is nonempty, we will have the product of variables $\mathbf{\delta A^+}$, $\mathbf{\delta A^-}$ with $\mathbf{x}$, which leads to a nonconvex mixed-integer quadratically constrained program (MIQCP). MIQCPs are computationally much more expensive than MILPs. Furthermore, most of the left-hand side parameters in optimization problems correspond to physical properties that cannot be changed easily. Therefore, OptiChat is designed such that the left-hand parameters are not recommended to be changed unless the user insists. We will describe how to achieve this in the next subsection.

\subsection{Implementation and LLM Prompt Engineering}
The LLM plays a central role in understanding the user's request and executing the optimization solver. OptiChat uses four prompting techniques to guarantee the reliability of performing the tasks. 
\begin{itemize}
    \item \textbf{Expert chain-of-thought (CoT)} We provide the LLM step-by-step instructions similar to how an optimization expert would approach each task.
    \item \textbf{Few-shot} Several simple examples of optimization problems are provided with expert answers.
    \item \textbf{Key-retrieve} We design this prompting technique tailored for OptiChat where we retrieve the keys of the parameter and the constraint names from the Pyomo code to improve robustness.
    \item \textbf{Sentiment} We prompt the LLM to understand the user's sentiment.
\end{itemize}
The application of these prompting techniques to the four major tasks of OptiChat is summarized in Table \ref{tab:prompts}.
\begin{table}[ht]
\caption{Prompting techniques used in the four main tasks of OptiChat: (T1)  Model analysis; (T2) Infeasibility diagnosis; (T3) Troubleshoot recommendation (T4) Interactive conversation.}\label{tab:prompts}
\scalebox{0.95}{
\begin{tabular}{lcccc}
\toprule
                            & Expert CoT & Few-shot & Key-retrieve & Sentiment \\
\midrule
T1             & $\checkmark$   & $\checkmark$        & $\checkmark$            &        \\
\midrule
T2     & $\checkmark$   & $\checkmark$        & $\checkmark$             &       \\
\midrule
T3 & $\checkmark$   & $\checkmark$        &              &        \\
\midrule
T4    &     &          &              & $\checkmark$          \\
\bottomrule
\end{tabular}}
\end{table}
In what follows, we describe how these prompting techniques are used in each task in combination with GPT-4's function-calling capability.
\subsubsection{Model analysis}
 We first use the Pyomo optimization tools to extract the model parameters and constraints from the code and embed their names in the prompt. In other words, we provide the LLM with the keys that are required to retrieve before instructing it to perform any action, namely the key-retrieve prompt. We empirically found that this key-retrieve prompt significantly reduces the rate of mismatching and misidentification when compared with summarizing the code directly. Second, we provide the steps an optimization expert would follow to describe the model (Expert CoT prompt), i.e., first provide an overview of the model, then summarize the input parameters, the decisions to be made, and the constraints. Third, few-shot demonstrations of the classic traveling salesman problem and the knapsack problem are used at each step of the Expert CoT.

\subsubsection{Infeasibility Diagnosis} When given an infeasible optimization, OptiChat uses Gurobi to find the IIS, returned as expressions of the constraints, with modeling parameters in numerical forms. OptiChat maps these constraints back to symbolic form to have a sensible natural language explanation. Expert CoT, few-shot, and key-retrieve prompting are used to generate reasonings for the infeasibility similar to the model description.

\subsubsection{Troubleshoot Recommendation} Parameters involved in the IIS are extracted using Pyomo modeling APIs. Recommendations are made to the user based on whether these parameters can be changed at low costs in the real world. Another criterion is trying to avoid changing the left-hand side parameters so that the model is kept to be an MILP. These considerations are conveyed to the LLM through expert CoT and few-shot prompting where several examples of parameters and their adjustability in the real world are given to the LLM, e.g., the capacity of a storage vessel can be expanded but the number of hours in a day cannot be changed.

\subsubsection{Interactive Conversation} Advanced LLM like GPT-4 \cite{OpenAI2023} provides chatbot APIs that enable the user to have interactive conversations. Besides answering the questions related to the model description and infeasibility analysis, one notable feature of OptiChat is that it can decide when to solve the problem with slack variables in \eqref{eq:MILPwithSlack} based on the conversations with the user. The user may agree or disagree with the initial troubleshooting recommendations provided by OptiChat in the previous step. Under certain cases, the user may want to try changing the parameters that OptiChat would not recommend. To handle this, we deploy a sentiment prompting technique. This does not strictly prohibit users from modifying them but gives rise to a warning about their consequences before the user's confirmation.
After the user confirms the parameters to change, a function calling API we implemented is executed to solve the optimization problem \eqref{eq:MILPwithSlack} where only the parameters deduced from the conversation with the user can be adjusted.

\section{Experiments}
In this section, we present a comprehensive overview of the outcomes achieved by OptiChat. 

We evaluate the effectiveness of our chatbot for understanding a wide variety of optimization models by performing a study on 8 inexperienced and 7 experienced users, who troubleshoot 20 and 38 infeasible instances, respectively. Through this real-world human study, we assess OptiChat's ease of use, response quality, and accuracy in interpreting the models and guiding the users to repair them.
The results show that both groups of participants highlight the effectiveness of OptiChat in understanding the optimization models and simplifying the troubleshooting process of realistic optimization models.

\subsubsection{Dataset preparation} 
To assess OptiChat, we select several feasible optimization models from the GAMS Model Library\footnote{\url{https://www.gams.com/latest/gamslib_ml/libhtml/index.html}}, the Pyomo Cookbook by the University of Notre Dame\footnote{\url{https://jckantor.github.io/ND-Pyomo-Cookbook}}, and a Resource Task Network model\footnote{\url{https://github.com/hdavid16/RTN-Demo}}. The selected problems correspond to a wide range of applications, including aircraft allocation with uncertain demand, production scheduling, production distribution and inventory, power generation scheduling, oil refining, bid evaluation, ship allocation, and military manpower planning.

Our dataset includes 63 infeasible instances generated by changing the original optimization problems in two ways: modifying a model parameter (e.g., minimum inventory, demand, maximum capacity) or adding constraints (e.g., maximum cost, minimum demand of a particular product).

\subsection{Infeasiblity Troubleshooting Assessment}
In this subsection, we assess the time to analyze and troubleshoot infeasible optimization models by experienced users and the strategies they use to repair such models. Our assessment is conducted using a survey with closed- and open-ended questions.

The group of experienced users repaired 38 infeasible optimization instances. The majority of the group reported three repairing strategies: (i) verifying the irreducible infeasible subset, (ii) activating/deactivating constraints, and (iii) relaxing constraints by adding slack variables. Despite the effectiveness of such strategies, it is noteworthy that the latter two strategies are prone to human errors and time-consuming depending on the size of the model. We note that such drawbacks can be circumvented by using HCI systems \cite{Greenberg1983}.

We breakdown the time to analyze and repair the optimization model into the following tasks:

\begin{itemize}
\item \textbf{Model analysis:} Corresponds to the time it takes to read the optimization model, i.e., the Pyomo script, and understand the model, e.g., identify the objective, constraints, decision variables, and parameters.
\item \textbf{Model description:} Pertains to the time it takes to write a detailed description of the optimization model.
\item \textbf{Infeasibility diagnosis:} Corresponds to the time it takes to diagnose the potential causes of infeasibility, e.g., conflicting constraints and/or parameters.
\item \textbf{Infeasibility repair:} Includes the time it takes to find a feasible solution once the potential causes of infeasibility have been detected.
\end{itemize}

\begin{figure*}[t]
     \centering
     \includegraphics[width=\textwidth,trim={0 1.95cm 0 0.1cm},clip]{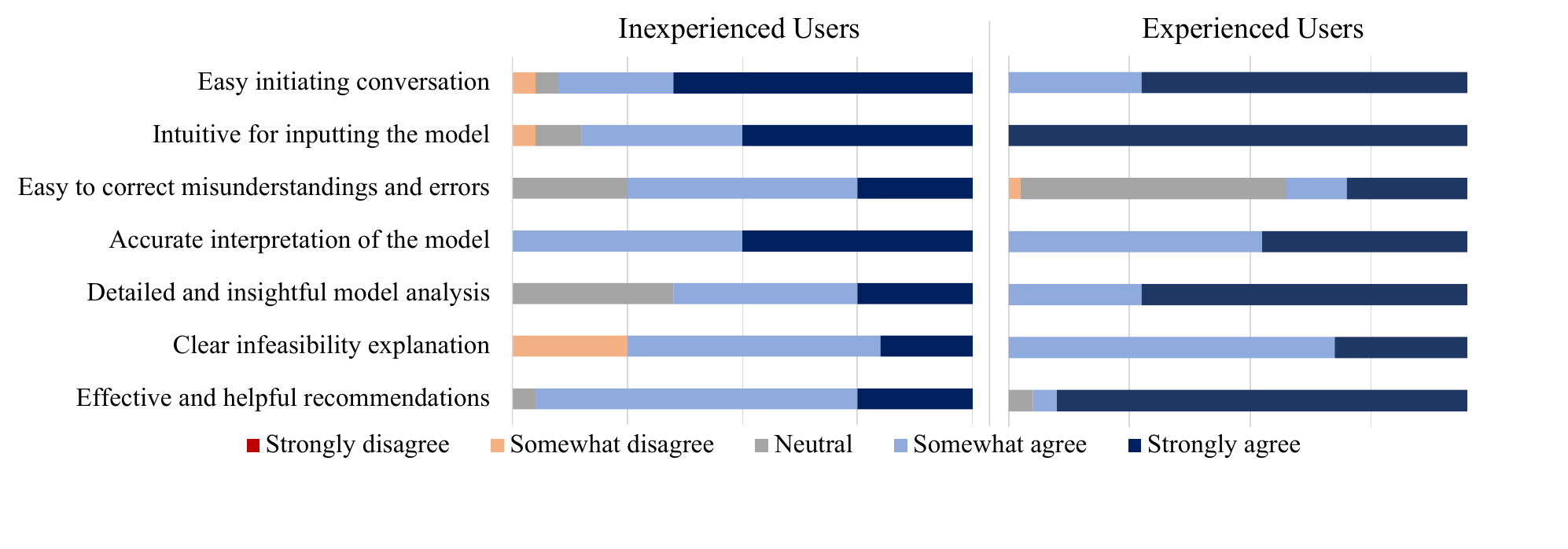}
    \caption{User study results: Likert graph of survey statements.}
    \label{fig:likert}
\end{figure*}

Most users spend 15-30 minutes analyzing the optimization model, as shown in Table~\ref{tab:time_manual}. We note that the completion times vary according to the familiarity of the user with the models. It is reasonable to assume that inexperienced users will require significantly more time to perform such tasks, highlighting the importance of the proposed natural-language-based system. Additionally, most optimization models in real-world applications are considered large-scale problems, i.e., problems with tens of thousands of variables and constraints. Clearly, the size of the optimization model could be, and generally is, detrimental to the time required to repair such problems, even for experienced users.

\begin{table}[h]
\centering
\caption{Infeasibility repair results: Percentage of users per time slot across tasks.}
\label{tab:time_manual}
\scalebox{0.925}{
\begin{tabular}{lcccc}
\toprule
                        & \multicolumn{4}{c}{Time (min)}                 \\
\midrule
Task                    & \textless{}5 & 5-15 & 15-30 & \textgreater{}30 \\
\midrule
Model analysis          & 7.9\%        & 10.5\% & 76.3\%  & 5.3\%              \\
Infeasibility diagnosis & 28.9\%       & 57.9\% & 10.5\%  & 2.6\%              \\
Infeasibility repair    & 52.6\%       & 39.5\% & 2.6\%   & 5.3\%              \\
\midrule
                        & 1-3          & 3-10 & 10-20 & \textgreater{}20 \\
\midrule
Model description       & 26.3\%       & 42.1\% & 5.3\%   & 26.3\%             \\
\bottomrule
\end{tabular}}
\end{table}

\subsection{OptiChat Assessment}
In this subsection, we assess different criteria to evaluate OptiChat's performance, usability, and effectiveness, similar to previous works that evaluate the coordination of human and ML models. We recruited two groups of users with and without experience in troubleshooting optimization models. We asked the participants to repair a number of randomly selected instances using OptiChat. Their utterances can include questions specific to the model, general optimization questions, and troubleshooting recommendations. Similar to the previous subsection, the assessment uses a survey with close- and open-ended questions, allowing us to assess the chatbot quantitatively and qualitatively, respectively.

\subsubsection{Quantitative Results}
The survey includes questions and statements with 1-5 Likert scale and open-ended questions to gather qualitative user feedback. The statements aim to evaluate the user experience, easiness of correcting chatbot errors or misunderstandings, accuracy and insightfulness of model interpretation, clarity of infeasibility explanation, and quality of troubleshooting recommendations.

Figure~\ref{fig:likert} shows the results of the quality assessment. For both groups, the modes of the responses across all the interactions lie in the \textit{somewhat agree} to \textit{strongly agree} categories except in the case of the statement about the easiness of correcting misunderstandings and errors during the interaction with the chatbot in the group of experienced users where the mode is \textit{neutral}.

At the end of the survey, the participants were required to report the chatbot's answers that they considered unsatisfactory due to misinterpretations of the questions or inaccurate answers, which allows us to compute the percentage of satisfactory answers with respect to the total. Additionally, the participants reported their success rate for troubleshooting the assigned infeasible optimization instances using OptiChat. Table~\ref{tab:rate_chatbot} shows the high accuracy rate for both metrics across the study groups.

\begin{table}[h]
\centering
\caption{OptiChat's accuracy results.}
\label{tab:rate_chatbot}
\scalebox{0.95}{
\begin{tabular}{lcc}
\toprule
\multirow{2}{*}{Study group}            & Satisfactory  & Troubleshooting  \\
                                        & answers        & success rate \\
\midrule
Inexperienced          & 90.93\%        & 88\%     \\
Experienced            & 87.20\%        & 96.77\%     \\
\bottomrule
\end{tabular}}
\end{table}

\subsubsection{Qualitative Results}
Based on the feedback provided in the survey, we provide representative quotes. Participants have highlighted the clarity of the explanations using easy-to-understand language provided by OptiChat, which can be helpful for users with different levels of optimization knowledge,

\begin{displayquote}
    \textit{It clearly explained the optimization problem in a simple and clear manner so that people without much optimization background could also understand the model. It also explains the meaning of different parameters and what it implies in the context of the considered model and real-life feasibility. I liked the feature of explaining complicated stuff with real-life example, not related to the considered problem.}---Participant 1.
\end{displayquote}

Participants also commented on the systematic understanding provided by OptiChat, which is achieved due to two prompt engineering techniques, namely, chain-of-thought and few-shot learning. Additionally, participants highlight the ability of OptiChat to repair the models after the initial troubleshooting recommendation,

\begin{displayquote}
    \textit{It is very helpful that they can provide the literal explanation on what to improve while also providing some systematic understanding of the each problem formulation. Also their recommendation on questions asking one parameter change has been always accurate in the first round of asking to make problem feasible.}---Participant 4.
\end{displayquote}

Overall, participants in both groups have expressed positive feedback regarding OptiChat, emphasizing the clarity of the explanations, structure and insightfulness of the responses, and timely responses. Additionally, participants have provided feedback that can guide future research directions, mainly referring to the language used while explaining infeasibilities. That is, we could adjust the language used to explain the models and infeasibilities depending on the user's level of expertise. 

\section{Conclusions}
In this manuscript, we propose OptiChat, an LLM-powered agent system that facilitates analyzing and troubleshooting infeasible optimization models using natural language-based interactions with non-expert users. Our agent relies on several prompting techniques to improve the robustness, including expert chain-of-thought, few-shot learning, and sentiment prompting. Additionally, to improve OptiChat's reliability, we propose a prompting technique named \textit{key-retrieve}, which allows extracting the keys corresponding to the parameter and constraint names directly from the optimization model script. The aforementioned prompting techniques are used in combination with GPT-4's function-calling capability to interact with state-of-the-art optimization solvers. The proposed chatbot is capable of troubleshooting linear and mixed-integer linear optimization problems, which can model a wide variety of applications in finance, logistics, and critical infrastructure planning and operation.

\end{document}